# KM3NeT : a large underwater neutrino telescope in the Mediterranean Sea

**Petros A. Rapidis**
Institute of Nuclear Physics, NCSR `Demokritos', Agia Paraskevi Attikis, Athens 15310, Greece

For the KM3NeT consortium

E-mail: rapidis@inp.demokritos.gr

**Abstract.** High energy neutrinos produced in astrophysical processes will allow for a new way of studying the universe. In order to detect the expected flux of high energy neutrinos from specific astrophysical sources, neutrino telescopes of a scale of a $km^3$ of water will be needed. A Northern Hemisphere detector is being proposed to be sited in a deep area of the Mediterranean Sea. This detector will provide complimentary sky coverage to the IceCube detector being built at the South Pole. The three neutrino telescope projects in the Mediterranean (ANTARES, NEMO and NESTOR) are partners in an effort to design, and build such a $km^3$ size neutrino telescope, the KM3NeT. The EU is funding a 3-year Design Study; the status of the Design Study is presented and some technical issues are discussed.

## 1. Introduction

Detecting high-energy neutrinos from astrophysical sources will be a major step towards a more complete understanding of the universe. These neutrinos can be detected by water/ice Cherenkov telescopes. Such detectors can contribute to the study of active galactic nuclei, supernova remnants, micro-quasars, gamma ray bursts, etc. In addition searches for neutrinos from the decay or annihilation of dark matter particles (WIMPs), magnetic monopoles and other exotic particles can be carried out [1]. Studies indicate [2] that such signatures will only be detectable by detectors of sizes of a $km^3$ of water or larger. Present existing telescopes [3], [4] are too small for this task.

The IceCube [5] telescope being built at the South Pole will have an instrumented volume of ice of a $km^3$. Neutrino telescopes are mostly downward looking detectors, and as a result the IceCube device will have a limited reach for sources in the southern sky. Thus the Galactic plane with a multitude of possible high energy neutrino sources, such as supernova remnants, microquasars, pulse wind nebulae, as well as several unassociated gamma-ray sources reported by the H.E.S.S. telescope[6] cannot be studied by IceCube.

To address this lack of sky coverage a neutrino telescope in the Mediterranean Sea has been proposed. The three pilot neutrino telescope projects in the Mediterranean (ANTARES [4], NEMO [7], and NESTOR [8]) are partners in an effort to design, and build such a $km^3$ size neutrino telescope, the KM3NeT [9]. The Mediterranean Sea offers some unique advantages for such a device: deep sites near the shore, clear waters, and periods of good weather needed for sea operations. The KM3NeT consortium consists of 37 Institutes from 10 European Countries [10]. The KM3NeT will be an interdisciplinary research infrastructure, serving as a deep water facility for associated sciences, like marine biology, oceanography, earth and environmental sciences, in addition to being a neutrino telescope.

## 2. Status of KM3NeT

The EU has initiated a 3-year design phase for KM3NeT which started in February 2006, and will conclude in 2010 with the publication of a Technical Design Report. A Preparatory Phase (PP) will follow, overlapping for one year with the Design Study. The PP will address political, governance, and financial issues of KM3NeT, including the site selection. The PP will also include prototyping work, in view of the start of the telescope construction in 2011. The overall cost of KM3NeT is estimated to be 220 -250 M€. KM3NeT is part of the ESFRI (European Strategic Forum on Research Infrastructures) roadmap [11] for future large scale infrastructures.

## 3. Detector Layout

The Design Study has as a goal the design of a detector with the highest sensitivity. Minimum requirements are an instrumented volume of at least 1 $km^3$, with angular resolution of about $0.1^o$ for neutrino energies above 10 TeV, sensitivity to all neutrino flavors, and a lower energy threshold of a few hundreds of GeV (and around 100 GeV for pointing sources). Neutrino Cherenkov telescopes are lattices of a large number of photo-detection units that detect the Cherenkov light. These units, the optical modules (OM), contain one or more photomultipliers (PMT) and are enclosed in a pressure resistant, waterproof glass sphere. The OMs used to date contain one large PMT (typically 10”). The OMs are arranged so as cover the instrumented volume. The specific choice of PMTs, their arrangement in the OM, as well as the layout of OMs in the instrumented volume are being studied with detailed MC simulations, in order to optimize the sensitivity of the detector.

Various OM configurations and detector layouts have been studied [12]. Examples of possible OMs can be seen in Fig. 1. These include OMs with a single PMT, double OMs, and OMs with many small PMTs inside. A number of detector layouts have been evaluated, with strings of vertically equidistant OMs arranged in cubic, ring, hexagonal, clustered or mixed layouts (Fig. 2). The instrumented volume has been kept constant to 1 $km^3$ and the total photocathode area was kept constant.

The effective area for neutrinos[13] as a function of energy is shown in Fig. 3 where the KM3NeT configuration-1 detector consists of 127 vertical strings each with 25 OMs arranged in a hexagonal pattern, 100 m horizontal spacing, 15 m vertical spacing between OMs and 3 large PMTs per OM, similar to the ANTARES configuration. The configuration-2 detector is a structure of 225 vertical strings each having 36 OMs, arranged in a tetragonal pattern with inter-line horizontal spacing of 95 m, vertical spacing between OMs 16.5 m, with each OM containing 21 3” PMTs. The effective area calculation includes full simulation of the neutrino interaction, muon propagation, Cherenkov light transmission, track reconstruction and event selection. Optical noise of 40 kHz due to $^{40}K$ decays in the sea water was added, as measured by NESTOR [14] and NEMO [15] in the respective sites. The higher effective area of the KM3NeT detector configurations compared to Ice-Cube can be explained in terms of the higher photocathode area and the better angular resolution of the water detector.

## 4. Detector Components and Procedures

The recovery and repair of faulty components from the deep sea is very difficult. Therefore, the reliability of all deep sea components is absolutely essential. Design choices being considered towards improved reliability include simplifying the design and reducing the number of connections that have to be mated in the sea. The possibility of transmitting all data-to-shore via optical fibre and reduction of the in-sea electronics using a photonics-based network [16] is also being studied. In this scenario, the OM signal is impressed with an optical reflector/modulator on an optical signal coming in from the shore. The modulated signal is transmitted to shore and time stamped at the shore end, thus necessitating no lasers or digitization electronics in the deep.

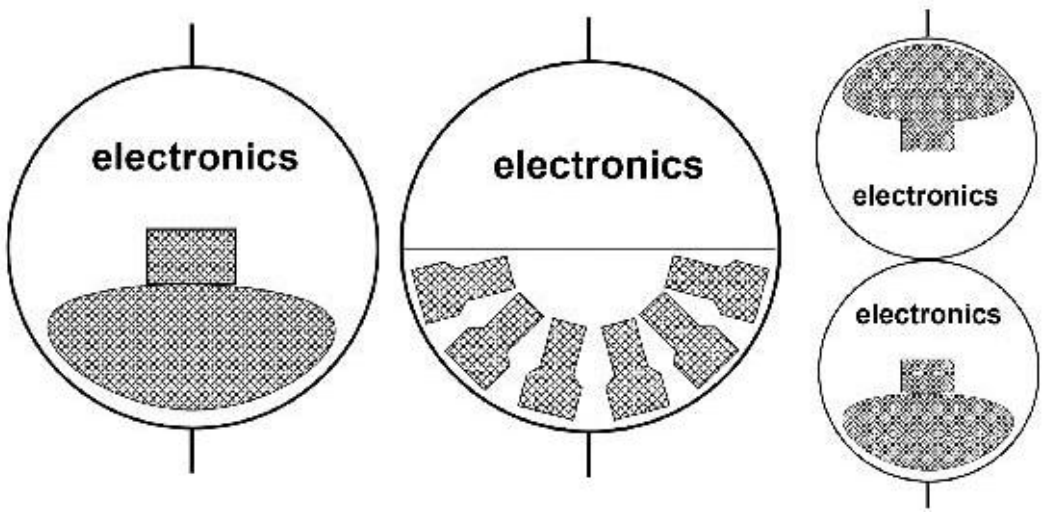

**Figure 1:** Various types of Optical Modules.

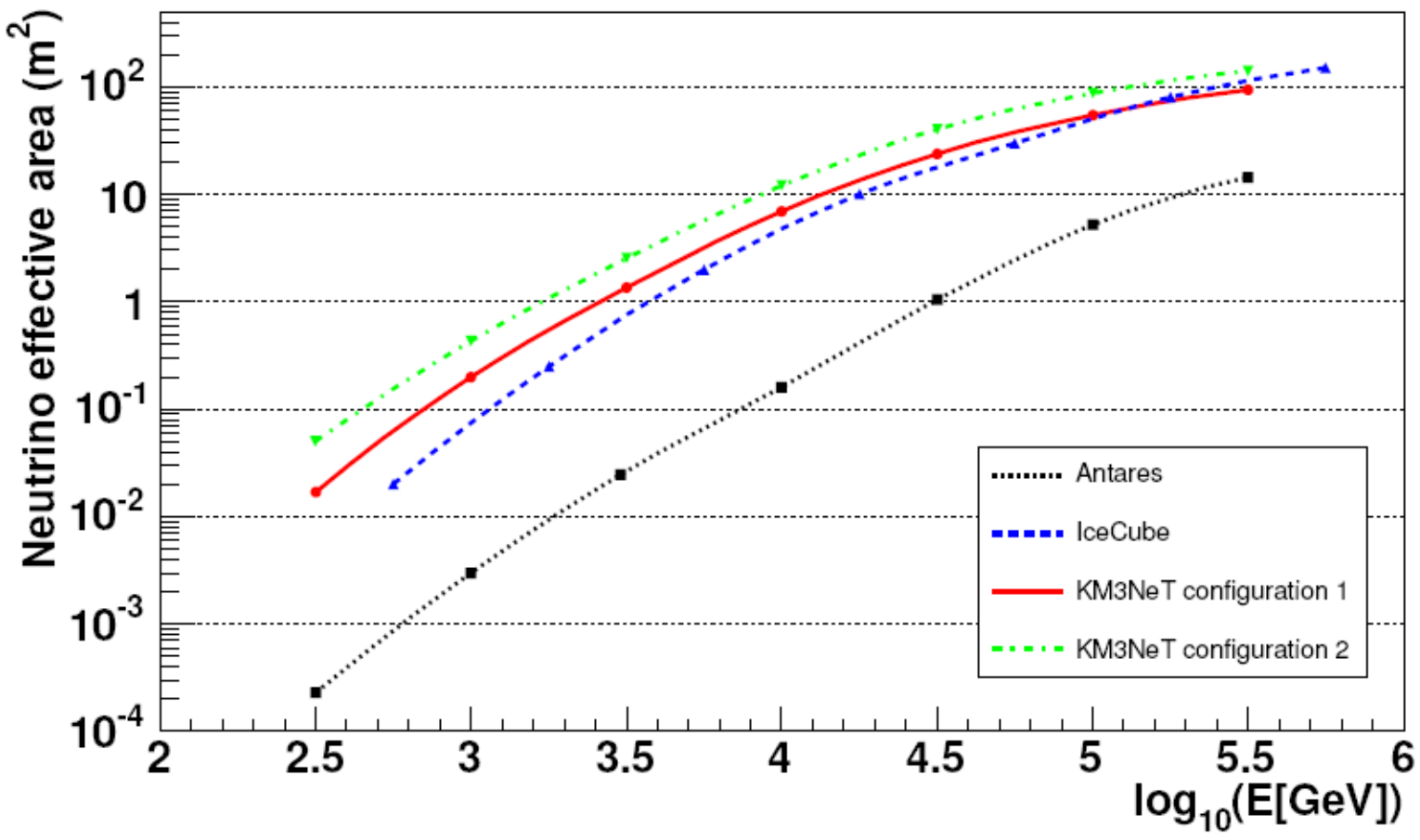

**Figure 3:** Neutrino effective areas vs. energy for two KM3NeT layouts, IceCube and ANTARES. The details of the KM3NeT configurations are described in the text.

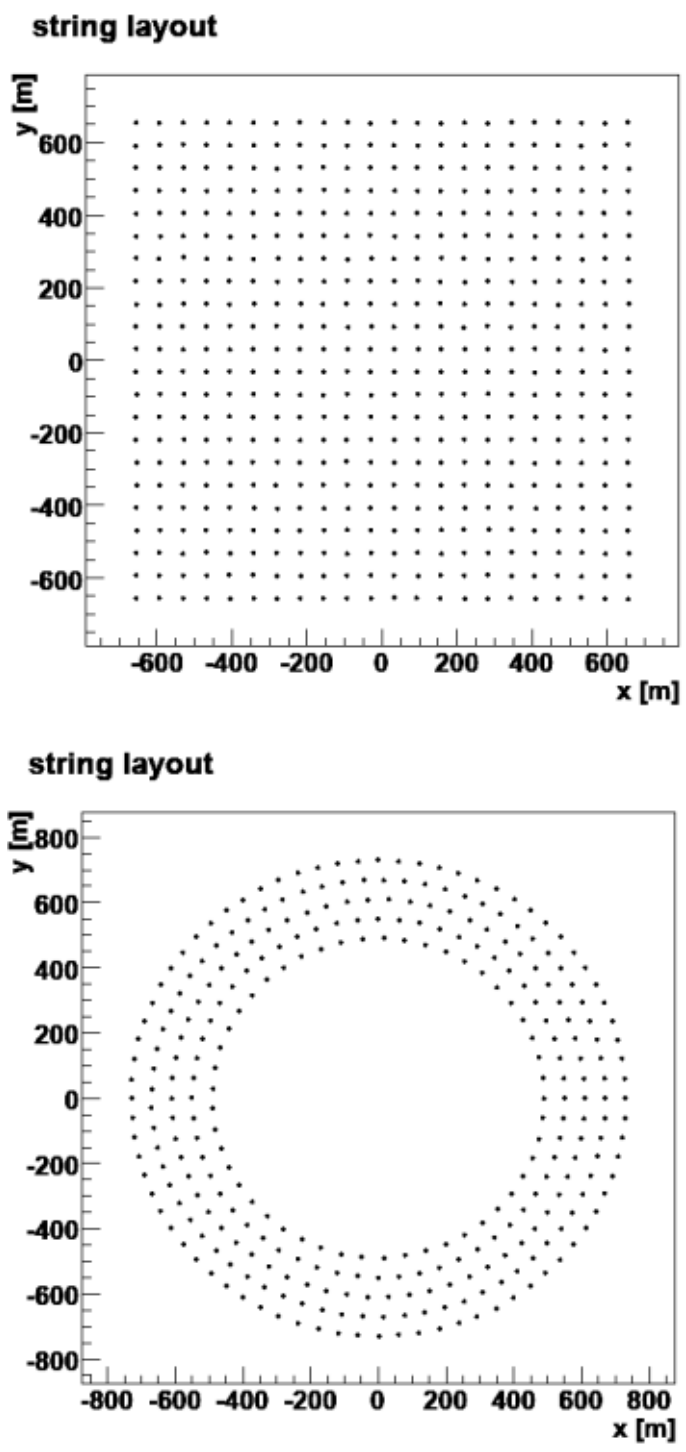

**Figure 2:** Cubic and ring detector layouts

Additional research and development is under way to evaluate the performance of OMs with many smaller PMTs[17]. These have the advantage of higher quantum efficiency, better single photon resolution, smaller transit time spread, and provide directionality which helps in reducing noise from bioluminescense and $^{40}$K decays in the sea water. Also under study are 10” PMTs with segmented photocathode area [18], as another way of gaining directionality, and crystal hybrid PMTs[19].

Calibration of angular resolution, absolute position and angular offset, that can be achieved by using a floating sea-top array [20], similar to IceTop is under investigation. The position of the sea-top detector can be determined via GPS, and the signal from extensive air showers can be used to calibrate the underwater detector. Three floating stations separated by distances of 20m equipped with 16m$^2$ of scintillator each will probably be sufficient for this scheme.

The sheer size of the detector implies that new ways of sea deployment will have to be developed. The ANTARES mode of deployment with separate operations for the line deployment and the cable connection with a submersible remotely operated vehicle may be time consuming and impractical for a detector the size of KM3NeT with around 250 lines. A possible solution could be the method proposed by the NEMO collaboration in which each line is ”rolled-up” in a container[21] equipped with a release mechanism which unfolds the line after the container reaches the sea floor. By connecting several of these compact containers together, one sea operation may suffice for the deployment of several lines. Such a deployment model reduces both the time the interconnection of lines and the number of connections that have to be mated in the sea which are a significant source of potential failures. The feasibility of such a deployment model has been succesfully demostrated by the NESTOR collaboration when they deployed a tower module without any underwater operation [22].

## 5. Associated Marine and Earth Sciences

The KM3NeT research facility will also serve the cause of marine and geophysical sciences. The existence of dedicated and permanent sea to shore connections allows the operation of long term real-time monitoring stations serving these disciplines. There is a separate working group within the Design Study, working on these issues. Besides monitoring instrumentation attached to the strings of the neutrino telescope additional dedicated associated sciences stations are envisaged.

## 6. References

[1] C. Spiering, 2005 Phys. Scripta, **T121**, 112

[2] A. Kappes *et al*., 2007 Astrophys.J, **656,** 870; M. D. Kistler and J. F. Beacom, 2006 Phys. Rev. D **74**, 063007

[3] The AMANDA detector at the South Pole, http://amanda.uci.edu

[4] The ANTARES detector in the Mediterranean Sea , http://antares.in2p3.fr

[5] The IceCube home page, http://icecube.wisc.edu/

[6] F. Aharonian *et al*., H.E.S.S. Coll., 2006 Astron.Astrophys**. 449 ,** 223; 2005 Astron.Astrophys**. 437**, L7

[7] The NEMO home page, http://nemoweb.lns.infn.it/

[8] The NESTOR home page, http://www.nestor.org.gr/

[9] The KM3NeT home page, http://www.km3net.org

[10] Cyprus, France, Germany, Greece, Ireland, Italy, The Netherlands, Romania, Spain, UK

[11] The ESFRI home page, http://cordis.europa.eu/esfri/home.html

[12] S. Kuch, Ph.D. Thesis, U of Erlangen, Germany, 2007, available from http://www.opus.ub.uni-erlangen.de/opus/volltexte/2007/692/

[13] The effective neutrino area, A, is defined by N=ΦA, where N is the number of detected neutrino events, and Φ is time integrated flux of incident neutrinos.

[14] G. Aggouras, *et al*., NESTOR Coll., 2005 Nucl. Inst. Meth. A **552**, 420

[15] T. Chiarusi, *et al*., NEMO Coll., Environmental parameters and water characteristics from deep-sea surveys at the NEMO sites, in *Proc. International Cosmic Ray Conference* (*ICRC05*), Pune, India, Aug. 2005

[16] P. Kooijman *et al*., Photonic readout of optical modules in neutrino telescopes, in *Proc. International Cosmic Ray Conference (ICRC07),* Merida, Mexico, July 2007

[17] P. Kooijman *et al*., Multi-PMT optical module for undersea neutrino telescopes, in *Proc. International Cosmic Ray Conference (ICRC07)*, Merida, Mexico, July 2007

[18] M. Taituti, Optical module for deep-sea neutrino telescopes,in *Proceedings for ICATPP07 Conference*, Como, 2007

[19] A. Braem *et a*l., 2007 Nucl. Inst. Meth. A **570**, 467

[20] A. Leisos, A Sea-Top Infrastructure for Calibrating an Underwater Neutrino Telescope, paper presented at *TeV Particle Astrophysics 2007*, Venice, Italy, 27-31 August 2007

[21] P. Sapienza, Status of the NEMO project, in *Proc. 20th European Cosmic Ray Symposium (ECRS 2006)*, Lisbon, 2006, e-Print: astro-ph/0611105; also contribution to this conference.

[22] E. Anassontzis *et al*., 2006 Nucl. Inst. Meth. A **567**, 538